\documentclass{ws-procs9x6adap} 

\usepackage{amsbsy}
\usepackage{epsfig}
\usepackage{amssymb}

\newcommand{\be}{\begin{equation}}
\newcommand{\ee}{\end{equation}}
\newcommand{\bea}{\begin{eqnarray}}
\newcommand{\eea}{\end{eqnarray}}
\newcommand{\bit}{\begin{itemize}}
\newcommand{\eit}{\end{itemize}}
\newcommand{\bra}{\langle}
\newcommand{\ket}{\rangle}

\newcommand{\im}{{\mathrm{Im}}}
\newcommand{\re}{{\mathrm{Re}}}

\newcommand{\cD}{{\mathcal {D}}}

\newcommand{\om}{\omega}

\newcommand{\tw}{\Gamma}
\newcommand{\aste}{{}^\ast\!}

\newcommand{\po}{p^{0}}
\newcommand{\qo}{q^{0}}

\newcommand{\gv}{\boldsymbol\gamma}
\newcommand{\gz}{\gamma^{0}}

\newcommand{\pv}{{\mathbf p}}
\newcommand{\kv}{{\mathbf k}}

\newcommand{\rv}{{\mathbf r}}

\newcommand{\pvuni}{\hat{{\mathbf p}}}
\newcommand{\kvuni}{\hat{{\mathbf k}}}

\newcommand{\rvuni}{\hat{{\mathbf r}}}

\newcommand{\puni}{\hat{p}}

\newcommand{\lmax}{\Lambda_{\rm max}}
\newcommand{\lmin}{\Lambda_{\rm min}}

\newcommand{\hs}[1]{\hspace*{#1}}

\newcommand{\al}{\alpha}

\newcommand{\bean}{\begin{eqnarray*}}
\newcommand{\eean}{\end{eqnarray*}}
\newcommand{\nn}{\nonumber}

\newcommand{\vecp}{{\mathbf p}}

\newcommand{\vecnul}{{\mathbf 0}}
\newcommand{\iop}{i0^{+}}

\begin{document}

\title{TRANSPORT COEFFICIENTS AND QUANTUM 
FIELDS\footnote{\uppercase{C}ombined invited talk by 
\uppercase{G}.~\uppercase{A}.\ and contributed poster by 
\uppercase{J}.~\uppercase{M}.~\uppercase{M}.~\uppercase{R}.\ presented at 
\uppercase{S}trong and \uppercase{E}lectroweak \uppercase{M}atter
(\uppercase{SEWM}2002), \uppercase{H}eidelberg, \uppercase{G}ermany, 2-5 
\uppercase{O}ctober 2002.} 
}

\author{Gert Aarts and Jose M.\ Martinez Resco}

\address{ Department of Physics, The Ohio State University\\
174 West 18th Avenue, Columbus, OH 43210, USA}

\maketitle 
\abstracts{ 
Various aspects of transport coefficients in quantum field theory are
reviewed. We describe recent progress in the calculation of transport
coefficients in hot gauge theories using Kubo formulas, paying attention
to the fulfillment of Ward identities. We comment on why the color
conductivity in hot QCD is much simpler to compute than the electrical
conductivity. The nonperturbative extraction of transport coefficients 
from lattice QCD calculations is briefly discussed.
}


\section{Introduction}

Transport coefficients characterize the response of a system in thermal
equilibrium to a weak perturbation associated with a conserved current.
Examples are the shear and bulk viscosities, electrical conductivity and
diffusion constants. Transport coefficients can be relevant in various
physical scenarios: in the formation and diffusion of large-scale magnetic
fields\cite{Giovannini:2001da}, in compact stars\cite{Alford:2002ng} and
in relativistic heavy-ion collisions\cite{Rischke:1998fq}.

The calculation of transport coefficients in thermal field theory through
Kubo formulas turns out to be highly nontrivial. The main problem is that
already at leading-logarithmic order an infinite class of diagrams, known
as ladder diagrams, has to be summed. This has favored the use of kinetic
theory where a few scattering amplitudes must be included in the collision
term. It is within the kinetic approach that it was first realized that
screening processes are necessary and sufficient to obtain finite
transport coefficients\cite{baym} and a complete leading-log calculation
for a variety of transport coefficients has appeared a few years
ago\cite{amy-tc}.

However, recent work\cite{valle} has contributed to establish an efficient
and easy way to compute transport coefficients to leading-log order using
the imaginary-time formalism of thermal field theory in a way that is
consistent with the Ward identity\cite{we}. In the following we focus on these
developments, describing the leading-log calculation of the electrical
conductivity. After that we comment on why the nonabelian color
conductivity turns out to be a much simpler quantity to compute. Finally
we discuss the use of lattice QCD as a nonperturbative approach to the
calculation of transport coefficients.


\section{Electrical conductivity}

In linear response theory transport coefficients are written as
equilibrium expectation values of commutators of currents. Indeed, the 
Kubo formula for the electrical conductivity in QED is
\be
 \label{eqkubo}
 \sigma = \lim_{\qo\to 0}\frac{1}{3}\,
 \frac{\partial}{\partial \qo}\im\,\Pi^{ii}_{R}(\qo,\vecnul),
\ee
where the retarded polarization tensor and electromagnetic current are
\be
 \Pi^{\mu\nu}_R(x-y) = i\theta(x^0-y^0)\bra[j^\mu(x),j^\nu(y)]\ket,
 \;\;\;\;\;\;
 j^\mu(x) = \bar\psi(x)\gamma^\mu\psi(x).
\ee
At weak coupling one might naively expect a one-loop calculation to be 
sufficient and such a calculation yields 
\be
 \label{eqsigmaone}
 \sigma = -\frac{2 e^2}{3} 
 \int_{\pv,\om} n_{F}'(\om)
 \left[ 
 \Delta^{R}_{+}(\om,\pv) \Delta^{A}_{+}(\om,\pv) 
 +
 \Delta^{R}_{-}(\om,\pv) \Delta^{A}_{-}(\om,\pv) 
 \right],
 \label{eqnaive}
\ee
where $\int_\pv = \int d^3p/(2\pi)^3$, $\int_\om = \int d\om/(2\pi)$ 
and $n_{F}$ is the Fermi distribution function. 
Here $\Delta^{R}_{\pm}$ denotes the retarded particle/anti-particle 
propagator and $\Delta^{A}_{\pm}$ the corresponding advanced one. 
These scalar propagators were introduced by decomposing the fermion 
propagator as 
\be
S(\om,\pv) = \Delta_{+}(\om,\pv)h_{+}(\pvuni) + 
\Delta_{-}(\om,\pv)h_{-}(\pvuni),
\ee 
with $h_{\pm}(\pvuni) = (\gamma^0\mp \gv\cdot\pvuni)/2$, $\pvuni=\pv/p$, 
and here and below we neglect the zero temperature electron mass for 
fermions with momentum $p=|\pv| \sim T$. In the free theory the scalar 
propagators 
 are
\be
 \Delta^{R}_{\pm,\rm free}(\om,\pv) = \frac{-1}{\om\mp p+i0^+}
 = \left[ \Delta^{A}_{\pm,\rm free}(\om,\vecp)\right]^{*}.
\ee
Since the free retarded (advanced) propagator has a pole at $\om = \pm p$
approaching the real axis from below (above), the products of the retarded
and advanced propagators as they appear in Eq.\ (\ref{eqnaive}) suffer
from so-called pinching poles: the integration over the energy variable in
Eq.\ (\ref{eqnaive}) is ill-defined and the naive result for the
conductivity is infinity!

\begin{figure}[t]
\centerline{
\includegraphics[height=1.5cm]{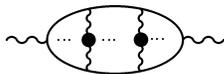}
}
\caption{Typical ladder diagram that contributes to the 
electrical conductivity at leading-logarithmic order. The side rails 
represent hard nearly on-shell fermions, dressed with a width, and the 
rungs are soft (HTL resummed) photons. 
}
\label{figladder}
\end{figure}

This particular result is of course well-known\cite{smilga,jeon-1} and 
has two major consequences. One has to  
\begin{itemize}
\item include a thermal width,
\item include an infinite series of ladder diagrams like the one in 
Fig.~\ref{figladder}.
\end{itemize}

The inclusion of the thermal width $\Gamma_\pv$ modifies the retarded and 
advanced single-particle propagators, which now read
\be
 \Delta^{R}_{\pm}(\om,\pv) = \frac{-1}{\om\mp p+i\tw_{\pv}/2}
 = \left[ \Delta^{A}_{\pm}(\om,\vecp)\right]^{*}.
\ee
As a result the pinching poles are screened (the distance between the 
poles is finite, namely $\Gamma_\pv$), which makes Eq.\ (\ref{eqnaive}) 
well-defined. Indeed, the products of the retarded and advanced 
propagators are now
\be
 \label{eqpp}
 \Delta^{R}_{\pm}(\om,\pv)\Delta^{A}_{\pm}(\om,\pv) =
 \frac{1}{(\om\mp p)^{2}+(\tw_{\pv}/2)^{2}}
 \longrightarrow \frac{2\pi}{\tw_{\pv}}\delta(\om\mp p),
\ee
where the last equation is valid in the limit of weak coupling.

The second consequence, the need to sum all ladders, is technically more
involved. For the shear viscosity in scalar field theory this feature has
been first recognized and implemented in full detail by Jeon\cite{jeon-1}.
His analysis has subsequently been confirmed and simplified by a number of
groups\cite{Wang:1999gv}. For gauge theories the problem was realized by
Lebedev and Smilga\cite{smilga} ten years ago, but no complete calculation
has been provided. Mottola and Bettencourt\cite{Mottola:sewm} have 
discussed the possibility of extracting the electrical conductivity from a 
consistent truncation of the Schwinger-Dyson hierarchy, but again without 
presenting a complete calculation.

\begin{figure}[t]
\centerline{
\includegraphics[scale=0.57]{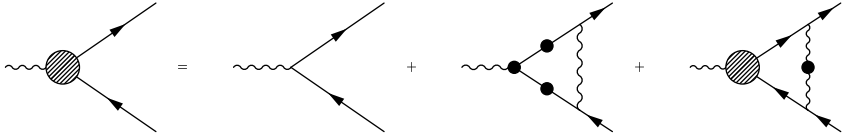}
}
\caption{Integral equation for the effective electron-photon vertex, 
$\Gamma^\mu = \gamma^\mu +\Gamma^\mu_{\rm HTL} + \Gamma^\mu_{\rm ladder}$. The 
HTL propagators and vertices are indicated with black blobs. The 
external photon has zero momentum and vanishing energy. Other lines 
represent hard nearly on-shell particles.
}
\label{figvertex}
\end{figure}

This unsatisfactory situation for gauge theories changed only recently
when Valle Basagoiti\cite{valle} used a concise method to sum ladder
diagrams, borrowing techniques from the condensed-matter literature, and
obtained equations for the shear viscosity and electrical conductivity to
leading-logarithmic order in (non)abelian gauge theories that are
equivalent to those obtained before using effective kinetic
theory\cite{amy-tc}. One way to sum all the ladders diagrams for the
electrical conductivity in QED is to introduce an effective
electron-photon vertex defined by an integral equation whose formal
solution is the geometric series summing all the rungs in the ladder.
Although the method presented by Valle Basagoiti solved in a simple way
the problem of summing the ladder series, his treatment missed one
important point since his integral equation was not consistent with the
Ward identity. Indeed, in order to fulfill the Ward identity an additional
diagram has to be included in the equation for the effective
vertex\cite{we} and the correct integral equation is depicted in
Fig.~\ref{figvertex} (the second diagram on the RHS is the new necessary
element).  However, although this extra diagram is essential to fulfill
the Ward identity, it does not contribute to the conductivity at
leading-log order\cite{we}. These results have been confirmed
recently\cite{Boyanovsky:2002te}.

Using the techniques previously mentioned\cite{valle} the complete
expression for the electrical conductivity at leading-log order can now 
be written in a way that closely resembles the one-loop result. It reads
\bea  
 \label{eqsigma}
 \sigma  = -\frac{2}{3}e^{2}\int_{\pv,\om}
 n_{F}'(\om) 
 \big[ \Delta^{R}_{+}(\om,\pv)\Delta^{A}_{+}(\om,\pv)
 \,\puni^{i}\,\re\, D_{+}^{i}(\om,\om;\pv) && \nn  \\ 
 - \Delta^{R}_{-}(\om,\pv)\Delta^{A}_{-}(\om,\pv) 
 \,\puni^{i}\,\re\, D_{-}^{i}(\om,\om;\pv)\big], &&
\eea
where we defined
\bea
 \label{eqDplus}
 D^{\mu}_{+}(\om+\qo,\om ; \pv) &\equiv& 
 \bar{u}_{\lambda}(\pvuni) \tw^{\mu}(\om+\qo+\iop,\om-\iop;\pv) 
 u_{\lambda}(\pvuni),  \\
 \label{eqDmin}
 D^{\mu}_{-}(\om+\qo,\om ; \pv) &\equiv& 
 \bar{v}_{\lambda}(\pvuni) \tw^{\mu}(\om+\qo+\iop,\om-\iop;\pv)
 v_{\lambda}(\pvuni).
\eea
$\tw^{\mu}$ represents the effective vertex and $u_{\lambda}$
($v_{\lambda}$) are spinors for the electron (positron) in a simultaneous 
chirality-helicity base. Since the conductivity is dominated by the 
pinching-pole contribution,  
out of the many different electron-photon vertices $\Gamma^i$ with 
real energies\cite{Carrington:1996rx} only one particular analytical
continuation contributes. 
Recalling Eq.~(\ref{eqpp}), it is convenient to define
\be \label{equu}
 \cD(p)\equiv \pm\puni^{i} \re\, D^{i}_{\pm}(\pm p,\pm p;\pv),
\ee
where we made use of rotational invariance and CP 
properties of the vertex, and the electrical conductivity 
is given by
\be
\label{eqsigmacon}
 \sigma=-\frac{4e^{2}}{3}\int_{\pv}n_{F}'(p)\frac{\cD(p)}{\tw_{\pv}}.
\ee
The 4 arises from electrons and positrons with either helicity that 
contribute in the same way to the conductivity.


\section{Ward identity}

The diagrammatic evaluation of transport coefficients has two main
ingredients: the inclusion of the thermal width which modifies the 
propagator and the summation of an infinite series of ladder diagrams 
which modifies the electron-photon vertex. 
In a gauge theory these two features have to be related, since the 
propagator and the vertex are connected through the Ward identity
\be 
 \label{wi}
 Q_{\mu}\Gamma^{\mu}(P+Q,P) = S^{-1}(P)-S^{-1}(P+Q). 
\ee
For the specific situation we are considering, the Ward identity reads
\be
 \qo\tw^{0}(\po+\qo+i0^{+},\po-i0^{+};\pv)
 = \qo\gz + \Sigma^{A}(\po,\pv)-\Sigma^{R}(\po+\qo,\pv),
\ee
in the special kinematical regime relevant for the conductivity, i.e.\ 
$\qo\to 0$ and $p^{0}\simeq \pm p$.
In terms of the quantity 
\be
 \label{eqdefD}
 {\mathfrak{D}}(\pv) \equiv \lim_{\qo\to 0} \qo
 D_\pm^0(\pm p+\qo, \pm p;\pv), 
\ee 
the Ward identity takes a particularly simple form
\be
 \label{eqWI}
 \mathfrak{D}(\pv) = i\tw_\pv.
\ee 
Since only the imaginary part of the self-energy is resummed, the RHS of 
the equation above is purely imaginary. From the definition in 
Eq.\ (\ref{eqdefD}) it is 
clear that the imaginary part of the vertex should diverge as $1/q^0$ when 
$q^0\to 0$. We want to verify that the Ward identity is indeed satisfied, 
by computing both sides of Eq.\ (\ref{eqWI}) independently.

\begin{figure}[ht]  
 \centerline{
 \includegraphics[height=1.5cm]{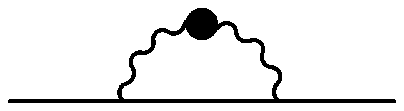}
 \includegraphics[height=1.5cm]{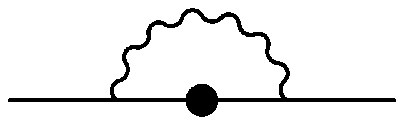}
 }
 \hs{3.3cm}(sp) \hspace{3.5cm} (sf) 
 \caption{Contributions to the thermal width of a hard on-shell fermion
 from a soft photon (sp) and a soft fermion (sf).}
 \label{figselfenergy}
\end{figure}

The thermal width $\tw_\pv$ of an on-shell electron with hard momentum
can be computed from the imaginary part of the self-energy. 
The leading terms with logarithmic sensitivity to the coupling constant 
arise from the diagrams shown in 
Fig.~\ref{figselfenergy}, with 
$\Gamma_\pv=\Gamma_\pv^{\rm (sp)}+\Gamma_\pv^{\rm (sf)}$.
The soft-photon contribution can be written as 
$\Gamma_\pv^{\rm (sp)} = \Gamma_\pv^{\rm (sp,lo)} +\Gamma_\pv^{\rm (sp,nlo)}$
for the first two terms with logarithmic sensitivity. 
The explicit expressions are\cite{blaizot,we} 
\bea
 \Gamma_\pv^{\rm (sp,lo)} &=& 2\al T\ln(m_D/\lmin), \\
 \Gamma_\pv^{\rm (sp,nlo)} &=& \frac{\alpha\, m_D^2\ln(1/e)}{2p}
 \left[-1+2n_{F}(p)+\frac{p}{6T} + 2p\,n_{F}'(p) \right],
\eea
where $\alpha=e^2/4\pi$ and $m_D = eT/\sqrt{3}$ is the Debye mass. 
The leading-log contribution from soft fermions is\cite{valle,we}
\be
 \label{eqsf}
 \Gamma^{\rm(sf,lo)}_{\pv} = \label{widthsofte}
 \frac{\alpha\,m_f^2\ln(1/e)}{p}\left[1+2n_{B}(p)\right],
\ee
with $m_f = eT/\sqrt{8}$ the fermionic thermal mass and $n_{B}$ the 
Bose distribution function. 
The technical reason why one needs to include the next ``logarithmic"  
order is that just doing a naive leading-log order calculation (i.e.\
keeping only $\Gamma_\pv^{\rm (sp,lo)}$) leads to an integral equation
without solution. We demonstrate this below. Furthermore, $\Gamma_\pv^{\rm
(sp,lo)}$ is actually ill-defined\cite{blaizot} and diverges
logarithmically; $\lmin$ is an ad-hoc infrared cut-off introduced to
regulate the divergence.  It is then clear that for the method to be
meaningful any dependence on this piece of the thermal width must
disappear in the end. The physical reason is that the electrical
conductivity is determined by large-angle Coulomb scattering along with 
pair annihilation and Compton scattering\cite{amy-tc}: this sets the 
relevant scale to be $e^{4}T\ln(T/m_{D})$. The leading term in the thermal 
width, however, arises from the exchange of ultrasoft quasistatic gauge 
bosons which corresponds to small-angle scattering. Processes which give 
the next ``logarithmic" order of the thermal width (as can be seen by cutting 
the diagrams) are precisely those related to the scale $e^{4}T\ln(T/m_{D})$; 
in the soft-fermion case directly through the thermal width
$\Gamma^{\rm(sf,lo)}_{\pv}$ and in the soft-photon case in a more
subtle way through the integration over the rung in the integral equation.
It is therefore understandable (of course, now that both the underlying
physics and the technical details are known) that subleading terms in the
thermal width must be included and that the dependence on the leading one 
in the end should disappear. We will see how this works in detail in the 
next section.
 
\begin{figure}[tb]
\centerline{
\includegraphics[scale=0.57]{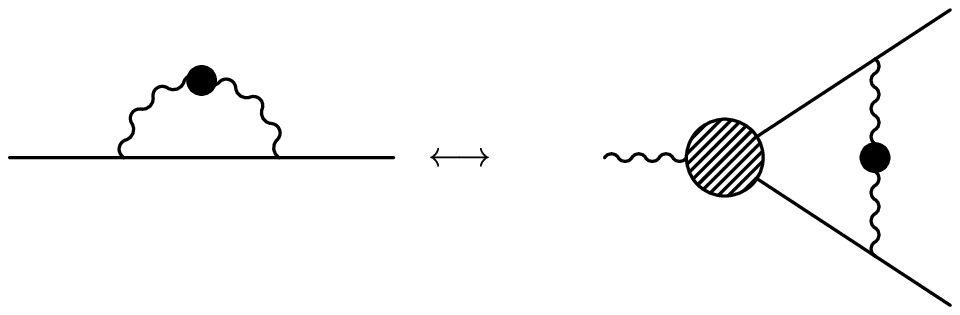}
\includegraphics[scale=0.57]{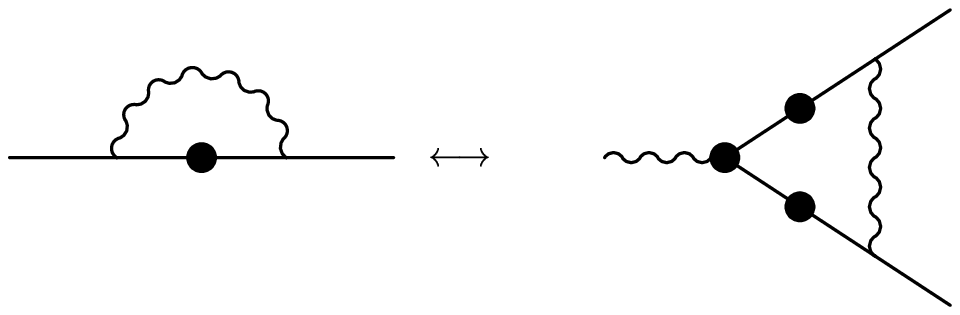}
}
\hs{1.1cm}$\Gamma_\pv^{\rm (sp)}$\hs{2.3cm}$\Gamma^0_{\rm ladder}$
\hs{1.5cm}$\Gamma_\pv^{\rm (sf)}$\hs{2.6cm}$\Gamma^0_{\rm HTL}$
\caption{The imaginary parts of the fermion self-energy and the 
two terms in the integral equation for the electron-photon vertex
are directly related to each other via the Ward identity. The self-energy 
with a soft photon (sp) corresponds in the effective vertex to the term 
that leads to the infinite series of ladder diagrams, the self-energy
with a soft fermion (sf) corresponds to a vertex correction with HTL 
vertex and propagators.
}
\label{figward}
\end{figure}

We now to turn to verify that our integral equation $\Gamma^0 = \gamma^0 +
\Gamma^0_{\rm HTL} + \Gamma^0_{\rm ladder}$ (see Fig.\ \ref{figvertex}) is
consistent with the Ward identity. It can be checked in detail\cite{we}, 
but here we will just argue that Eq.\ (\ref{wi}) makes it natural to expect 
that the inclusion of a contribution to the self-energy with soft fermion 
lines must go with a corresponding contribution to the vertex. The  
correspondence is shown in Fig.~\ref{figward}. 
The contribution of $\Gamma^0_{\rm ladder}$ to $\mathfrak{D}(\pv)$ is
precisely the soft-photon contribution to the thermal width, while the new
diagram $\Gamma^0_{\rm HTL}$ is precisely the required piece to obtain the
soft-fermion part of the $\tw_{\pv}$.  Therefore, the inclusion of the
thermal width and the summation of ladder diagrams are in fact closely
related and one cannot do one without the other. This is relevant for 
computations beyond leading-log: any additional diagram that contributes 
to the thermal width should be reflected in corresponding new 
contributions to the integral equation for the effective vertex and vice 
versa.


\section{Leading-log result}

To obtain the final result for the electrical conductivity the integral
equation for the spatial part of the effective electron-photon vertex
still has to be solved. Because only the real part of the effective vertex
is needed, see Eq.~(\ref{equu}), the calculation simplifies considerably
since the real part of the new diagram $\Gamma_{\rm HTL}^{i}$ is
subleading and therefore does not contribute\cite{we}.

The integral equation for the vertex $\cD(p)$ can be written as\cite{we}
\bea  \label{ie}
 &&\cD(p)  =1 + \nn  \frac{\al}{2 p^{2}} 
 \int_{\lmin}^{\lmax}\!\!\!\!\!\! dk\,k \int_{-k}^{k}\frac{d\om}{2\pi}
 \left[n_{B}(\om)+n_{F}(p+\om)\right] 
 \left\{ \pvuni\cdot\rvuni\frac{\cD(r)}{\tw_{\rv}}\Big|_{z=z_{0}} 
 \right\}  \\  
 && \times 
 \left[\aste\rho_{T}(\om,k)
 \frac{k^{2}-\om^{2}}{k^{2}}\left[(\om+2p)^{2}+k^{2}\right]
 +\aste\rho_{L}(\omega,k)\left[(\om+2p)^{2}-k^{2}\right] \right].
 \label{ie0}
\eea
Here $\aste\rho_{T/L}(\om,k)$ are the spectral densities for the soft
transverse/longi\-tu\-di\-nal photons in the rung, $\rv=\pv+\kv$,
$z=\kvuni\cdot\pvuni$, and $z_0 = \om/k + (\om^2-k^2)/(2pk)$. $\lmax$ is
an upper cut off introduced to be consistent with the condition that the
photon carries soft momentum; at leading-log accuracy\cite{braaten} it can
be taken to be $T$. Save for the factor within braces, the integral is
precisely the soft-photon contribution $\tw^{(\rm{sp})}_{\pv}$ to the
thermal width.

At this moment we can demonstrate the technical reason why keeping just
the leading term $\sim e^2T$ to the thermal width is inconsistent. In the
vertex equation the on-shell particles carry hard momentum $p$ and the
collective HTL modes carry soft momentum $k$. This scale separation allows to
expand the integrand in powers of $k/p$. At (naive) leading 
order, the term within the braces is just $\cD(p)/\tw_{\pv}$ and can be 
taken out of the integral. The integral equation then reduces to
$\cD(p) = 1+ \cD(p)$ which has no solution! 

We therefore proceed keeping subleading contributions to the width. In order
to show that any dependence on the scale $\sim e^2T$ drops out, we write
\be
\chi(p) =  \frac{\cD(p)}{\tw_\pv}, \;\;\;\;\;\;
\sigma =-\frac{4e^{2}}{3}\int_{\pv} n_{F}'(p)\chi(p).
\ee 
Expanding, as before, in powers of $k/p$ leads now to a differential 
equation for $\chi$, which with leading-log accuracy reads
\bea
 1 &=& \frac{\alpha\, m_{f}^{2}\ln(1/e)}{p} \left[1+2n_{B}(p)\right] 
 \chi(p) 
 + \frac{\alpha\, m_{D}^{2}\ln(1/e)}{p} \frac{T}{p}
 \nn\\ && \times
 \left[ \chi(p) -
 \left(1-\frac{p}{2T}\,[1-2n_{F}(p)]\right) p\,\chi'(p)
 -\frac{1}{2}p^{2}\chi''(p) \right].
\label{eqdiff1}
\eea
The only scale present in this equation is $e^4T\ln(1/e)$, as it should 
be. There is no dependence on $\lmin$. The parametrical dependence of 
the conductivity can be made explicit by writing
\be
 \chi(p) = \frac{T}{\alpha\,m_D^2 \ln(1/e)}\,\phi(p/T),
\ee
such that
\be \sigma = C\frac{T}{e^{2}\ln(1/e)}, \;\;\;\;\;\;\;\;\;\;\;\;\;\;
 C = \frac{2}{\pi}\int^{\infty}_{0} dy\, y^2\frac{1}{\cosh^2(y/2)}\phi(y),
\label{eqC}
\ee
and the dimensionless function $\phi(y)$ obeys the differential equation
\be
1 = \left[ \frac{3\coth(y/2)}{8y}+\frac{1}{y^2}\right]\phi(y)
+ \left[\frac{1}{2}\tanh(y/2)-\frac{1}{y}\right]\phi'(y)
-\frac{1}{2}\,\phi''(y).
\label{eqdiff}
\ee
The integral in Eq.\ (\ref{eqC}) is dominated by hard ``momentum" 
$y=p/T$. In the limit of large $y$ the differential equation 
(\ref{eqdiff}) simplifies considerably and is solved by 
the particular solution $\phi(y)=8y/7$. 
Using this approximate result yields 
\be
C_{\rm approx} = \frac{288}{7\pi}\zeta(3) \approx 15.7424,
\ee
which is close to the exact result $C=15.6964$, obtained by 
AMY\cite{amy-tc} using a variational approach.


\section{Soft fermions}

As can be seen above, the relevant inverse time scale for the electrical
conductivity is $\sim e^{4}T\ln(1/e)$. This scale enters the integral
equation determining the effective vertex $\chi(p)$, see Eq.\
(\ref{eqdiff1}), and arises partly from the integration over the
soft-photon rung and partly through the explicit appearance of the
soft-fermion contribution to the thermal width (the first term on the RHS
of Eq.\ (\ref{eqdiff1})).
Since in the diagrammatic calculation the soft-fermion contribution to the
thermal width appears explicitly in the equation for the conductivity and
since the imaginary part of the self-energy is directly related to scattering
processes (as can be seen by cutting the diagrams), we expect that there is a direct
relation between it and the inverse relaxation time from the corresponding
scattering processes included in the collision term in the kinetic approach. 
The reason why we do not expect so
with the soft-photon contribution is that it is ill-defined and the
corresponding process contributes in a more subtle way to the integral equation,
as explained above.

\begin{figure}[b]
\centerline{
\includegraphics[height=3cm]{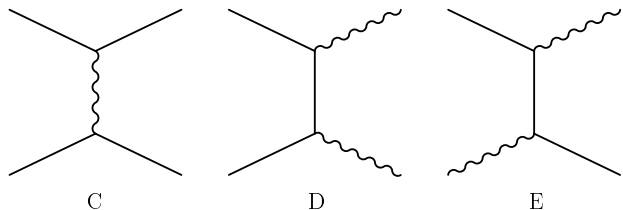} }
\caption{Scattering diagrams
that contribute to the electrical conductivity at leading-logarithmic
order in hot QED in kinetic theory (time runs horizontally). }
\label{figscatt}
\end{figure}

The scattering processes that contribute to the electrical conductivity at
leading-log order in kinetic theory\cite{amy-tc} are shown in
Fig.~\ref{figscatt}: large-angle Coulomb scattering (diagram C),
pair creation/annihilation (D) and Compton scattering (E) (throughout this
section we follow the notation of AMY\cite{amy-tc}).

Since, as explained above, we are interested here in scattering processes 
where a soft fermion is exchanged, we restrict ourselves in the following to 
diagrams D and E. For two-to-two scattering processes the collision term 
reads 
\bea
\nn &&
 C[f](\pv) = \frac{1}{2}\int_{\kv, \kv', \pv'}
 \frac{\left| {\mathcal M}(\pv,\kv;\pv',\kv')\right|^2}{2^4pp'kk'}
 (2\pi)^4\delta^4(P+K-P'-K')
\\ && \nn
 \times \left\{ f(\pv) f(\kv)[1\pm f(\pv')][1\pm f(\kv')]
 - f(\pv') f(\kv')[1\pm f(\pv)][1\pm f(\kv)] \right\},
\\ &&
\eea
where the $+/-$ signs refers to bosons/fermions.
If the distribution function for the incoming fermion with
momentum $\pv$ is perturbed slightly away from equilibrium, $f(\pv) =
n_F(\pv) + \delta f(\pv)$, while all other distribution functions (bosonic
and fermionic) are kept in equilibrium (i.e.\ using a relaxation-time
approximation), the collision term can be written as
\be
C[f](\pv) = \frac{1}{\tau_\pv}\delta f(\pv),
\ee
with the inverse relaxation time given by
\be
\label{eqtau}
\frac{1}{\tau_\pv} = \frac{1}{2(4\pi)^4p^2} \int^T_{eT} dq
\int_{-q}^{q} d\om \int_0^\infty dk \int_0^{2\pi} d\phi \,
|{\mathcal M}|^2\, [\mbox{stat}.],
\ee
and $[\mbox{stat}.]$ is what remains of the statistical factors.
Here $\om = p'-p$ and $q=|\pv'-\pv|$ are the energy and momentum of
the exchanged fermion and the integral over $q$ is cut off in the infrared 
by
the expected Debye scale. Note that in the leading-log calculation
using kinetic theory this is the only place where medium effects appear.
To leading-log accuracy, one finds\cite{amy-tc}
\be
|{\mathcal M}|^2_{\rm D} =|{\mathcal M}|^2_{\rm E} =
\frac{16e^4 pk}{q^2}(1-\cos\phi),
\ee
and
\bea
&&\!\!\!\!\! [\mbox{stat}.]_{\rm D} =
n_F(k)[1+n_B(p)][1+n_B(k)] + n_B(p)n_B(k)[1-n_F(k)],
\\
&&\!\!\!\!\! [\mbox{stat}.]_{\rm E} =
n_B(k)[1+n_B(p)][1-n_F(k)] + n_B(p)n_F(k)[1+n_B(k)].
\eea
The integrals in Eq.\ (\ref{eqtau}) can now be performed. To leading-log
accuracy the relaxation rates associated with D and E are identical and
the total relaxation rate corresponding to processes where a fermion is
exchanged is, in the leading-log approximation,
\be
 \label{eqtausf}
 \frac{1}{\tau_\pv} \Big|_{\mbox{fermion exchanged}} = e^4\ln(1/e)
 \frac{T^2}{32\pi p} \left[1+2n_B(p)\right].
\ee
This is exactly the value of the soft-fermion contribution to the thermal
width, see Eq.\ (\ref{eqsf}), as expected.


\section{Color vs.\ electrical conductivity}

It is instructive to compare the diagrammatic computation of the
electrical conductivity with that of the color conductivity to 
leading-logarithmic order in hot QCD.
The color conductivity appears in the effective theory describing the
nonperturbative dynamics of ultrasoft modes of nonabelian gauge
fields\cite{Bodeker:1998hm}. 
It was first computed within kinetic theory\cite{selikhov,asy} and it
has also been obtained from a simplified ladder summation\cite{yo}. 
Although there is, as far as we know, no gauge
invariant definition of this quantity, one can compute it 
diagrammatically using the nonabelian generalization of the Kubo formula 
(compare with Eq.\ (\ref{eqkubo})),
\be
\sigma_c = \lim_{q^0\to 0} \frac{1}{3(N_c^2-1)}
\frac{\partial}{\partial q^0} \im\, \Pi_{ii R}^{aa}(q^0,\vecnul).
\ee
In principle one could think that the color conductivity is a more
complicated quantity to compute than the electrical conductivity due to
its nonabelian character. In reality the opposite is true. The reason is
that the nonabelian nature of the interactions allows small-angle
scattering to randomize the current by just changing the color charge of
the current carriers\cite{asy}. This means that in this case it is
sufficient to do a ``real'' leading-log calculation, i.e.\ it is enough to
keep the leading-order term of the thermal width $\sim g^2T\ln(1/g)$. In a
nonabelian theory the infrared logarithmic divergent behavior of this
width is not a problem because there is natural mechanism to regulate it,
the magnetic mass $m_g$.

\begin{figure}[t]
\centerline{\includegraphics[scale=0.57]{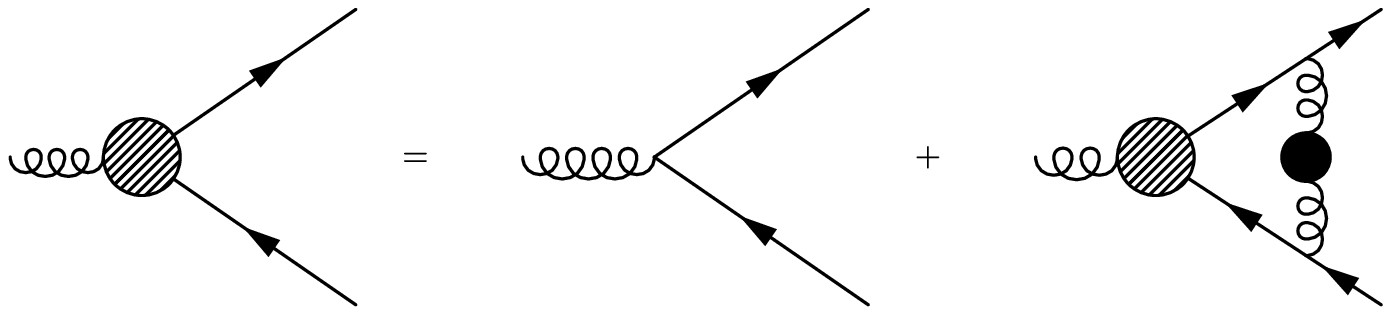}}
\centerline{\includegraphics[scale=0.57]{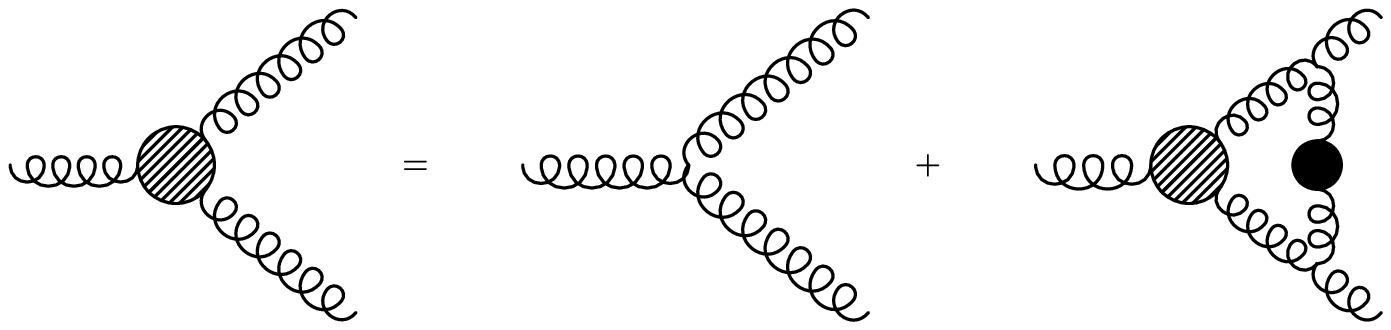}}
\caption{Integral equations for the effective quark-gluon vertex and 
three-gluon vertex, needed for the color conductivity at 
leading-logarithmic order. 
}
\label{figvertexcc}
\end{figure}

We now outline the diagrammatic calculation of the leading-log order of 
the color conductivity, following\cite{yo}, but using the exact and 
simplest way of carrying out the ladder summation. 
The calculation of the color conductivity goes along the same lines as the 
electrical conductivity. Since gluons are self-interacting there are 
both a quark and a gluon contribution. The integral equations are depicted
in Fig.~\ref{figvertexcc}. 
We find (compare with Eq.~(\ref{eqsigmacon}))
\be 
 \label{cc}
 \sigma_{c} = 
 -\frac{4g^{2}}{3}\frac{N_{f}}{2} 
 \int_{\pv}n_{F}'(p)\frac{\cD_{\rm q}(p)}{\tw_{\pv}^{\rm q}} 
 -\frac{4g^{2}}{3}\frac{N_{c}}{2}
 \int_{\pv}n_{B}'(p)\frac{\cD_{\rm g}(p)}{\tw_{\pv}^{\rm g}},
\ee
where $N_{f}$ is the number of flavors.
The quark-gluon effective vertex $\cD_{\rm q}(p)$ is defined similarly to 
the electron-photon one, see Eq.\ (\ref{equu}). 
For the gluonic vertex we note that since the gluons on the side rails 
carry hard momentum the longitudinal contributions are exponentially 
suppressed and the gluon propagators are proportional to the transverse 
projector $P^T$. The gluon scalar function $\cD_{\rm g}(p)$ is defined 
from the three-gluon effective vertex $\tw^{ijk}_{abc}$ after performing 
the analytical continuation, putting the hard gluons on the side rails 
on-shell, taking the energy $\qo$ of the external gluon to zero and 
contracting with transverse projectors, as
\be
 2f^{abc} p^{i} P^{T}_{jk}(\pvuni)\, \cD_{\rm g}(p) \equiv 
 P^{T}_{kk'}(\pvuni)\, \tw^{ij'k'}_{abc}\, P^{T}_{jj'}(\pvuni).
\ee
As in the case of the electrical conductivity the $\cD$'s represent the 
full vertex. The bare vertices correspond to $\cD_{\rm q}(p) = \cD_{\rm 
g}(p) = 1$.
The leading-order thermal width of quarks resp.\ gluons is\cite{pisarski}
\be
 \tw_{\pv}^{\rm q} = 
 \frac{N_{c}^{2}-1}{2N_{c}}\,2\al_{s}T\ln(m_{D}/m_{g}), 
 \;\;\;\;\;\;
 \tw_{\pv}^{\rm g} = N_{c}\,2\al_{s}T\ln(m_{D}/m_{g}),
\ee
with $m_{g}\sim g^{2}T$ the magnetic mass and therefore 
$\ln(m_{D}/m_{g})\sim\ln(1/g)$. 

It is now straightforward to adapt the calculation of the electrical
conductivity to the problem at hand: the quark contribution to the color
conductivity can easily be obtained just by inserting the correct group
factors. The integral equation for the effective quark-gluon
vertex is as in Eq.~(\ref{ie}), with the substitutions: $\al\to
-\al_{s}/(2N_{c})$ and $\lmin\to m_{g}$. At leading order in the 
expansion in $k/p$ the term within
braces is just $\cD_{\rm q}(p)/\tw_{\pv}^{\rm q}$. It is independent of
the integration variables and can be taken out of the integral. As we
mentioned for the electrical conductivity, the remaining integral is just
the thermal width, save for group factors in this case. The integral
equation reduces therefore to an algebraic equation
\be
 \cD_{\rm q}(p) = 1 - \frac{1}{2N_c} \times 
 \frac{\cD_{\rm q}(p)}{\tw_{\pv}^{\rm q}} \times 
 \frac{2N_c}{N_c^2-1} \tw_{\pv}^{\rm q}
 \;\;\;\;\Longrightarrow\;\;\;\;
 \cD_{\rm q}(p) = 1 - \frac{1}{N_c^2}.
\ee
The calculation of the gluon contribution is quite similar and 
again the integral equation can be solved algebraically: the 
solution is $\cD_{\rm g}(p) = 2$.

Since neither the effective vertices nor the thermal widths depend 
on momentum, the final result for the color conductivity from Eq.\ 
(\ref{cc}) is
\be
\sigma_c = \frac{g^2T^2}{9}\frac{N_f}{2}\frac{\cD_{\rm q}}{\Gamma^{\rm q}}
+ \frac{g^2T^2}{9}N_c\frac{\cD_{\rm g}}{\Gamma^{\rm g}}.
\ee
The ladder summation for the color conductivity is, as we see,  
substantially simpler than for other transport coefficients in hot gauge 
theories.


\section{Transport coefficients from the lattice}

Both kinetic and field theory allow to compute transport coefficients at
high temperature, where the coupling constant is small. However, it would also be
interesting to be able to compute transport coefficients at 
lower temperatures, where the coupling constant is no 
longer (very) small, which is relevant for heavy-ion collisions. 
In this section we discuss briefly the
prospects\cite{aarts,aartslattice} of extracting transport coefficients at
high temperature nonperturbatively from lattice QCD. As shown in Eq.\
(\ref{eqkubo}) for the electrical conductivity, transport coefficients can
defined from the slope of a spectral function $\rho(\om,\vecnul)$ at
vanishing energy ($\rho$ equals twice the imaginary part of the retarded
correlator). Spectral functions can be related to euclidean-time
correlators through a dispersion relation so that
\be
 \label{eqkernel}
 G_E(\tau,\pv )=\int_{0}^{\infty} \frac{d\om}{2\pi}\,
 K(\tau,\om)\rho(\om,\pv),
\ee
with the kernel $K(\tau,\om)  = n_B(\om) e^{-\om\tau} + [1+n_B(\om)]
e^{\om\tau}$. In the context of transport coefficients Eq.\
(\ref{eqkernel}) has been employed first by Karsch and
Wyld\cite{Karsch:1986cq} and more recently by Nakamura
et.~al.\cite{Nakamura:1996na} This approach consists of
three steps: 

\begin{itemize}
\item[$i)$] compute $G_E(\tau)$ (at zero momentum) numerically on
the lattice,
\item[$ii)$] reconstruct $\rho(\om)$ using either an
ansatz\cite{Karsch:1986cq} (old-fashioned approach) or the Maximal Entropy
Method\cite{Asakawa:2000tr} (modern approach), 
\item[$iii)$] extract the
transport coefficient from the slope at vanishing energy.
\end{itemize}

Recently we have analyzed what can be expected in the context of the shear
viscosity in scalar and nonabelian gauge theories at very high
temperature\cite{aarts} (the results can be adapted easily to other
transport coefficients such as the electrical conductivity). We found that
in a weakly-coupled field theory at high temperature the spectral function
has a characteristic shape. In particular, there is a bump at very small
energies which has its origin in the pinching singularities discussed
above. The height of this bump at small energies is $\sim 1$ in units of
the temperature.

The interesting question is how the spectral weight at small energies
manifests itself in the euclidean correlator. For small energies $\om \ll
T$ the kernel can be expanded as $K(\tau,\om) \simeq 2T/\om + {\mathcal
O}(\om/T)$. Since all the $\tau$ dependence resides in the subdominant
terms, the region relevant for transport coefficients contributes a
single, constant term to the euclidean correlator: $G_E(\tau) \sim \int
d\om\, \rho(\om)/\om$. We find therefore that although euclidean
correlators are sensitive to spectral weight at small energies in
integrated form, they are, in weakly coupled theories, remarkably
insensitive to further details of the spectral function in this region
and, therefore, also to transport coefficients.

The findings about the
small-energy region turn out to be rather generic\cite{aartslattice}.
They may therefore be relevant for recent attempts to reconstruct spectral
functions at finite temperature in the high-temperature deconfined phase of 
QCD using the Maximal Entropy Method\cite{Karsch:2001uw,Asakawa:2002xj}.


\section{Summary}

We have reviewed several aspects of the diagrammatic calculation of
transport coefficients in hot gauge theories to leading-log order. 
We focused in particular on recent progress within the imaginary-time
formalism to sum the ladder series in an efficient way, on the importance
of the Ward identity relating self-energy and vertex corrections, and on
similarities and differences between the color and electrical
conductivity. Finally, the prospects of extracting nonperturbative values
for transport coefficients using lattice QCD were briefly mentioned.


\section*{Acknowledgments}

\noindent
It is a pleasure to thank the organizers for a stimulating Workshop in a 
familiar (for one of us) environment.
Discussions with E.\ Braaten, U.\ Heinz, G.~D.\ Moore, and M.\ A.~Valle 
Basagoiti are gratefully acknowledged.

G.~A.\ is supported by the U.~S.\ Department of Energy under Contract No.\
DE-FG02-01ER41190.
J.~M.~M.~R.\ is supported by a Postdoctoral Fellowship from the Basque
Government and in part by the Spanish Science Ministry under Grants
AEN99-0315 and FPA 2002-02037 and by the University of the Basque Country
under Grant 063.310-EB187/98.


\end{document}